\documentclass{emulateapj}

\shorttitle{A wide-angle tail radio galaxy at $z=0.53$}
\shortauthors{A.~\oklop\ et al.}

\def\f#1   {Fig.~\ref{#1}}
\def\s#1   {Sec.~\ref{#1}}
\def\tab#1   {Tab.~\ref{#1}}
\def\t#1   {Tab.~\ref{#1}}

\def\comm#1   {{\tt (COMMENT: #1) }}

\def\kms{~km~s$^{\mathrm{-1}}$}

\def\sqdeg            {$\Box^{\circ}$}
\def\sqdegs           {$\Box^{\circ}$}

\def\msol              {$\mathrm{M}_{\odot}$}

\def\wh                {W~Hz$^{-1}$}

\def\kms{~km~s$^{\mathrm{-1}}$}

\def\oklop             {Oklop\v{c}i\'{c}}
\def\smo               {Smol\v{c}i\'{c}}

\slugcomment{  }

\begin{document}

\title{Identifying dynamically young galaxy groups via wide-angle tail
  galaxies:\\ A case study in the COSMOS field at
  $z=0.53$\altaffilmark{0}}

\author{A.~Oklop\v{c}i\'{c}\altaffilmark{1,2},
        V.~Smol\v{c}i\'{c}\altaffilmark{2,3,4}, 
        S.~Giodini\altaffilmark{5},
        G.~Zamorani\altaffilmark{6},
        L.~B$\hat{i}$rzan\altaffilmark{7}, 
        E.~Schinnerer\altaffilmark{8}, 
        C.~L.~Carilli\altaffilmark{9},
        A.~Finoguenov\altaffilmark{5},
        S.~Lilly\altaffilmark{10}, 
        A.~Koekemoer\altaffilmark{11},
	N.~Z.~Scoville\altaffilmark{2}
        }
\altaffiltext{0}{Based on observations with the National
Radio Astronomy Observatory which is a facility of the National Science
Foundation operated under cooperative agreement by Associated Universities,
Inc. }
\altaffiltext{1}{University of Zagreb, Physics Department,
  Bijeni\v{c}ka cesta 32, 10002  Zagreb, Croatia}
\altaffiltext{2}{ California Institute of Technology, MC 105-24, 1200 East
California Boulevard, Pasadena, CA 91125 }
\altaffiltext{3}{ESO ALMA COFUND Fellow, European Southern Observatory, Karl-Schwarzschild-Strasse 2, 
85748 Garching b. Muenchen, Germany}
\altaffiltext{4}{Argelander Institut for Astronomy, Auf dem H\"{u}gel 71, Bonn, 53121, Germany}
\altaffiltext{5}{Max Planck Institut f\"{u}r Extraterrestrische Physik, D-85478 Garching, Germany}
\altaffiltext{6}{INAF - Osservatorio Astronomico di Bologna, via Ranzani 1,
  40127, Bologna, Italy}
\altaffiltext{7}{Leiden Observatory, Leiden University, Oort Gebouw, P. O. Box 9513, 2300 RA Leiden, The Netherlands}
\altaffiltext{8}{ Max Planck Institut f\"ur Astronomie, K\"onigstuhl 17,
  Heidelberg, D-69117, Germany }
\altaffiltext{9}{National Radio Astronomy Observatory, P.O. Box 0, Socorro,
  NM 87801-0387 }
\altaffiltext{10}{Eidgen\"{o}ssische Technische Hochschule - Z\"{u}rich, CH - 8093 Z\"{u}rich, Switzerland}   
\altaffiltext{11}{Space Telescope Science Institute, 3700 San Martin Drive, Baltimore, MD 21218}

\begin{abstract}
We present an analysis of a wide-angle tail (WAT) radio galaxy located
in a galaxy group in the COSMOS field at a redshift of $z=0.53$
(hereafter CWAT-02). We find that the host galaxy of CWAT-02 is the
brightest galaxy in the group, although it does not coincide with the
center of mass of the system. Estimating a) the velocity of
CWAT-02, relative to the intra-cluster medium (ICM), and b) the
line-of-sight peculiar velocity of CWAT-02's host galaxy, relative to
the average velocity of the group, we find that both values are higher
than those expected for a dominant galaxy in a relaxed system. This
suggests that CWAT-02's host group is dynamically young and likely in
the process of an ongoing group merger. Our results are  consistent
with previous findings showing that the presence of a wide-angle tail
galaxy in a galaxy group or cluster can be used as an indicator of
dynamically young non-relaxed systems. Taking the unrelaxed state of
CWAT-02's host group into account, we discuss the impact of radio-AGN
heating from CWAT-02 onto its environment, in the context of the missing
baryon problem in galaxy groups. Our analysis strengthens recent
results suggesting that radio-AGN heating may be powerful enough to
expel baryons from galaxy groups.
\end{abstract}


\section {Introduction}
\label{sec:intro}

Wide-angle tail (WAT) radio galaxies are radio galaxies whose jets are bent
forming a wide C shape.  They are usually found in dense environments,
such as galaxy clusters and groups. It is believed that the
characteristic morphology of WATs is a result of ram pressure being
exerted on the radio jets, due to the relative motion of the host
galaxy with respect to the intra-cluster medium (ICM; e.g.\ Begelman,
Rees \& Blandford 1979).  WATs are generally associated with brightest
cluster galaxies (Owen $\&$ Rudnick 1976) and they are found to move with
peculiar velocities of a few hundreds \kms\ relative to the cluster
center. Often, such velocities cannot explain the shape of the radio
jets (e.g. Eilek et al. 1984). Therefore, an invoked scenario that can
explain the bending of the jets is one in which the ICM gas has
streaming flows driven by cluster/subcluster mergers. As a
consequence, the ICM gas may have an associated bulk velocity relative
to the potential and the galaxies (Klamer et al. 2004).

This picture is supported by observations which report the presence of
WAT galaxies in connection with other indicators of a recent cluster
merger, such as X--ray substructure (Burns et al. 1994), the
elongation of the X--ray emission along the line that bisects the WAT
(Gomez et al. 1997), and a significant offset ($\sim$ 100 kpc) of the
WAT from the X-ray centroid (Sakelliou \& Merrifield 2000). These
findings suggest that WATs are efficient tracers of dynamically young
clusters (see e.g.\ \smo\ et al.\ 2007; S07 hereafter).  This feature,
together with the ability to detect powerful radio galaxies with short
exposure at high redshift, where dimming affects the optical and X-ray
emission, makes WATs unique probes of clustering in the high redshift
universe (Blanton et al. 2003).

The interest in radio--galaxies has lately been renewed because of the
major role their AGN feedback may play in massive galaxy formation and
evolution (Croton et al.\ 2006; Bower et al.\ 2006; Best et al.\ 2006;
\smo\ et al.\ 2008), as well as heating of the gas in galaxy
clusters/groups. In terms of the latter, radio galaxies have been
proposed  to solve the 'cooling flow problem' in galaxy groups and
clusters (see Fabian 1994 for a review) and explain the lack of
baryons in galaxy groups (Bower et al.\ 2008; Short et al.\ 2008;
Giodini et al. 2009; 2010).

In recent cosmological simulations gas is removed from within the
cluster as a consequence of the mechanical heating by radio outflows
of the central AGN. This scenario successfully reproduces the observed
X-ray luminosity--temperature (L$_{X}$--T) relation and the halo gas
fractions (Bower et al.\ 2008; Short et al. 2008).  Only recently,
this idea has been observationally supported by Giodini et
al.\ (2010). Comparing the mechanical output energy of radio--AGN, in
a sample of COSMOS galaxy groups, with the group's binding energy,
they show that the mechanical removal of intra--cluster gas in galaxy
groups by radio--AGN heating may be energetically feasible. In
  particular, if WATs are efficient proxies for dynamically young,
  i.e.\ merging group environments, they then shed light onto the
  recent assembly history of the system. This is important in terms of
  the missing baryon problem on group scales as the binding energy in
  the final system is higher than that in the merging constituents,
  while the average power of radio--AGN outflows is expected to remain
  comparable \citet{smo08}. This highlights the importance of studying in detail the properties of
peculiar radio--galaxies within galaxy groups and their interaction with the
surrounding environment.

Here we present a multi-wavelength study of a wide-angle tail galaxy
at $z=0.53$ in the COSMOS field (CWAT-02 hereafter) and its
environment. This system is particularly interesting as it is at
intermediate redshift. Furthermore, CWAT-02 is a radio galaxy with a
one-sided jet, and its host group has already been studied by Giodini
et al.\ (2009; 2010) in the context of the missing baryon problem on
galaxy group scales.

We report magnitudes in the AB system, adopt $H_0=70,\, \Omega_M=0.3,
\Omega_\Lambda = 0.7$, and define the radio synchrotron spectrum as
$F_{\nu} \varpropto \nu^{-\alpha}$, where $F_{\nu}$ is the flux density
at frequency $\nu$ and $\alpha$ is the spectral index.

\section{Data}
The data used in this paper have been obtained from the panchromatic
(X-ray to radio) COSMOS 2\sqdeg\ survey (Scoville et al. 2007a), the 
largest mosaic obtained to date with the Advanced Camera for Surveys 
on HST (Koekemoer et al. 2007).  In the UV to IR wavelengths the field 
has been observed in $\sim30$ (broad and narrow) photometric bands 
(with GALEX, Subaru, CFHT, UKIRT and Spitzer; see Capak et al.\ 2007 
for details). The large number of bands allows a very accurate 
determination of photometric redshifts (Ilbert et al.\ 2009; Salvato et 
al.\ 2009) yielding a dispersion ($\sigma_{\delta z/\left(1+z_\mathrm{spec}\right)}$; 
where $\delta z=z_\mathrm{spec}-z_\mathrm{phot}$) of 0.7$\%$ at
$i_{AB}^+<22.5$, and 3.34$\%$ at $i_{AB}^+>22.5$.  The spectroscopic
data used for this work are drawn from the SDSS (York et al.\ 2000)
and zCOSMOS (Lilly et al.\ 2007, 2009) surveys.

 Within the VLA-COSMOS Large project (Schinnerer et al. 2007) the full
 2\sqdeg\ field was observed at 1.4~GHz with the NRAO Very Large Array (VLA)
 in A- and C- configurations. Additional A-array observations have
 been obtained for the inner 1\sqdeg\ (Schinnerer et al.\ 2010)
 yielding a final rms of $~\sim8~\mu$Jy/beam in the central part of
 the field at a resolution of $1.5"\times1.4"$.  The rms in the area
 around CWAT-02 is 10.5~$\mu$Jy/beam.

The COSMOS field has been observed at 324~MHz (90~cm) with the VLA in
A-array (November 2008) for a total of 24~hours (for a detailed
description see \smo\ et al., in prep). The full 2\sqdegs\ field has been
covered within one single pointing. The final map, used here, has been
generated using the AIPS task IMAGR and a weighting scheme
intermediate between natural and uniform
(i.e. $\mathrm{ROBUST}=0$). The reached $rms$ (fairly uniform over the
full 2\sqdegs ) and resolution are $\sim0.5$~mJy/beam and
$6.4''\times5.4''$, respectively. The $rms$ in the area of CWAT-02 is
$436~\mu$Jy/beam.

X-ray observations of the COSMOS field have been performed with both
XMM-Newton (1.5 Ms covering 2\sqdeg ; Hasinger et al.\ 2007) and Chandra 
(1.8 Ms in the inner 1\sqdeg ; Elvis et al.\ 2009). Here we make use of the galaxy
cluster catalog described in detail in Finoguenov et al.\ (2007; in
prep.). Extended source detection was based on a wavelet
analysis technique performed on the composite XMM-Newton and Chandra mosaic,
after background and point-source subtraction.  Each X-ray cluster
candidate was further independently verified via an optical galaxy
cluster search algorithm (making use of both the COSMOS photometric
and spectroscopic redshifts). The final catalog contains $\sim200$
galaxy clusters; the host group of CWAT-02 is identified as group \#35,
and its IAU designation is CXGG100049+0149.3.

\section{Properties of the wide-angle tail galaxy CWAT-02}

\subsection{Optical/IR}
\label{sec:opthost}

The host galaxy of CWAT-02 is located at $RA=150.2066\ (10:00:49.59)$
and $DEC=1.8233\ (+01:49:23.85)$. It is a bright ($M_I=-23.04$)
spheroid-like red (rest-frame $U-B=0.96$) galaxy with a spectroscopic
(SDSS DR7; \citealt{dr7}) redshift of $z=0.5302 \pm 0.0002$.  The
stellar mass of the galaxy, as estimated by \citet{smo08} by fitting
the global (from 3500 \AA\ to 2.5~$\mu$m) spectral energy distribution
with the \citet{bc03} stellar population synthesis models and a
\citet{chabrier03} IMF, is $3.5\times10^{11}$~\msol . This is
consistent with the highest mass galaxies known, which are often found
in dense environments and usually associated with central
brightest group galaxies (e.g.\ \citealt{best05}; von der Linden et
al. 2007). The host galaxy has not been detected with Spitzer/MIPS at
$24,70,160\ \mu$m suggesting a low amount of dust and insignificant
star formation. These properties are consistent with those of
red sequence passive galaxies.

\subsection{Radio}
\label{sec:radio}

\subsubsection{The structure and luminosity of CWAT-02}

\begin{figure*}
\center{
\includegraphics[bb = 50 370 540 720, width=1.0\columnwidth]{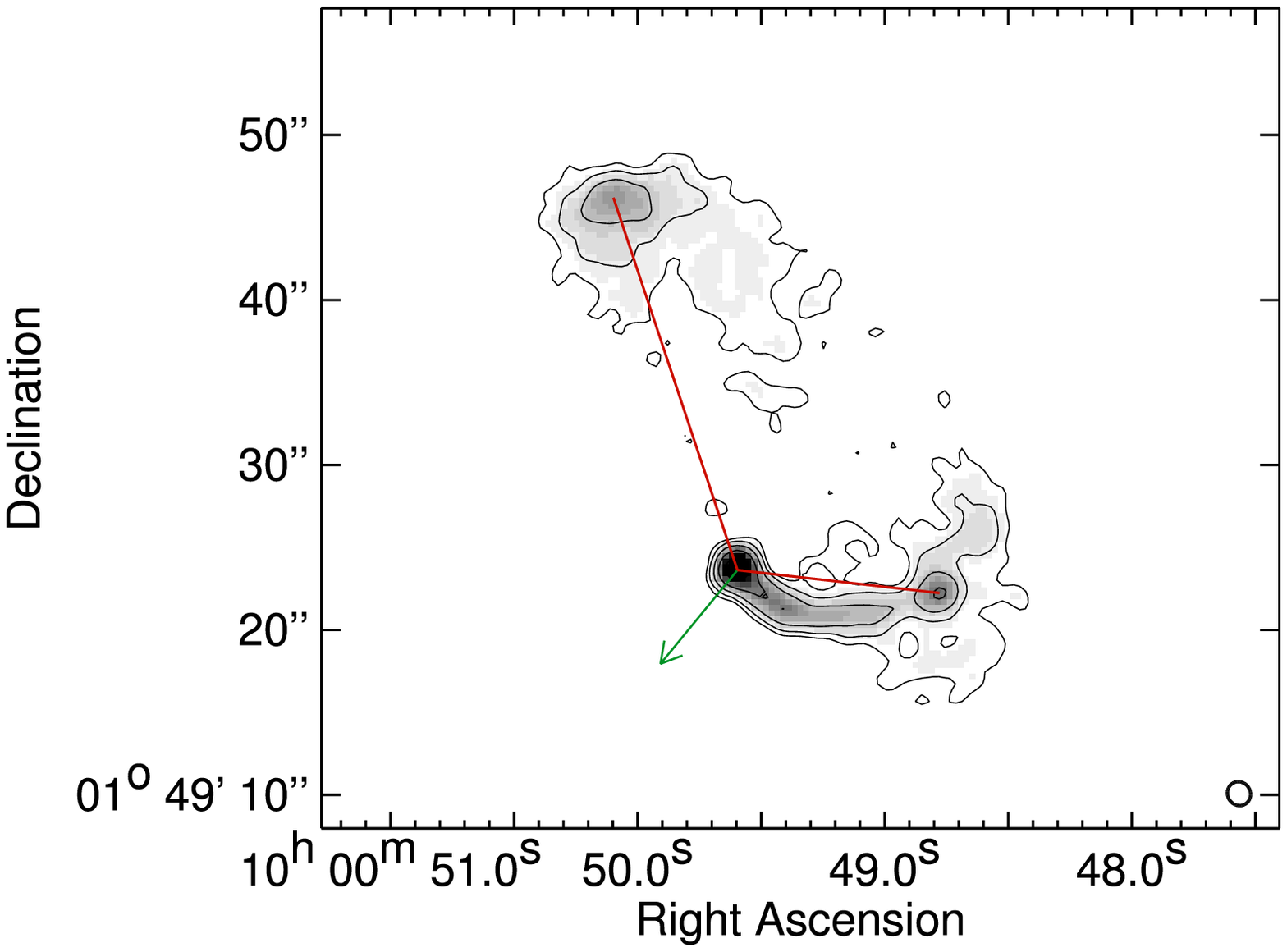}  
\includegraphics[bb = 50 370 540 720, width=1.0\columnwidth]{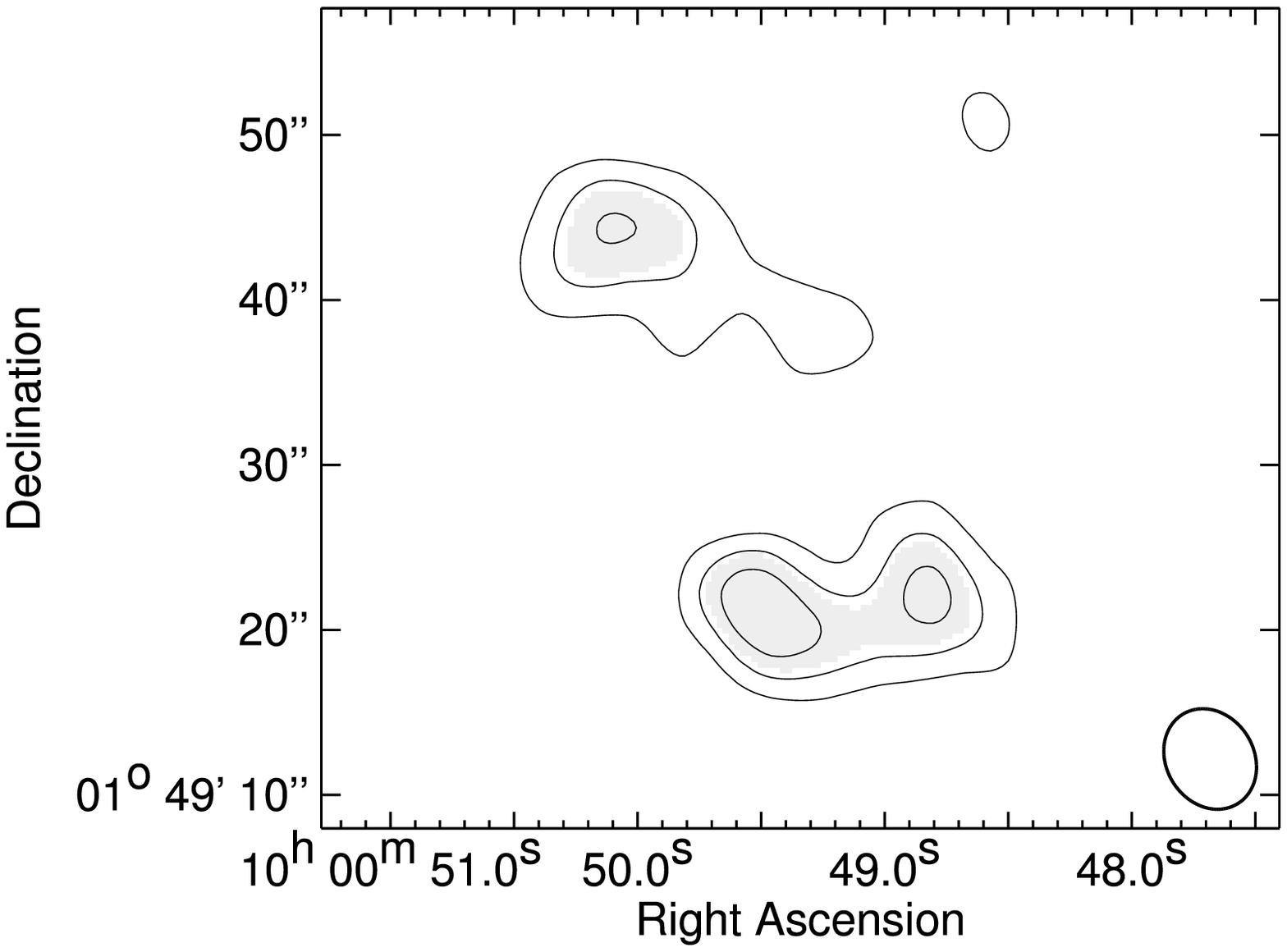}
\caption{ Left panel: 1.4 GHz radio image of CWAT-02. The contour
  levels are at $2^n\cdot\sigma$, where $n=2,3,4...$, and $\sigma=10.5
  \ \mu$Jy$/$beam.  The clean beam ($1.5'' \times 1.4''$) is shown in
  the lower right corner.  The arrow indicates the direction of the
  movement of CWAT-02 (relative to the ICM), in the plane of the sky.
  The lines connect the central radio core with the centers of the lobes,
  thus forming the bending angle of $\sim115^\circ$.
  Right panel: 324 MHz radio map of CWAT-02 with contours
  overlaid. The contour levels are in steps of $2\sigma$, starting at
  $3\sigma$ ($\sigma=0.436 \ $mJy$/$beam).  The resolution is $6.44''
  \times 5.38''$, and the clean beam is shown in the lower right
  corner.
\label{fig:radio}}
}
\end{figure*}

The radio morphology of CWAT-02, shown in \f{fig:radio} , is C-shaped,
consistent with the morphology of wide-angle tail galaxies.  The
central radio peak coincides with the above described elliptical
galaxy at $z=0.5302 \pm 0.0002$.  From the central radio core a very
luminous jet extends in the south-west direction out to a distance of
$\sim 20''$ ($\sim 120$ kpc at $z=0.53$). It ends in a lobe that
extends another $12''$ ($\sim 80$ kpc) in NW-SE direction (in the
plane of the sky). The counter-jet is not detected within the
VLA-COSMOS sensitivity, however, the lobe $\sim 25''$ ($\sim 155$ kpc
at $z=0.53$) away from the central core is clearly visible. Prominent
hot-spots are associated with both lobes. The asymmetry of the
jet/counter-jet luminosities that likely arises from Doppler boosting
suggests a small inclination of the jets to the line-of-sight (see
next section).  We estimate the direction of motion of CWAT-02
(relative to the ICM) toward SE (in the plane of the sky) by bisecting
the bending angle ($\sim115^\circ$) formed by the lobes (see
\f{fig:radio} ).

In \f{fig:radio} \ we also show the 324 MHz map of CWAT-02. Due to the
factor of $\sim4$ lower resolution and significantly decreased sensitivity 
($\sigma=0.436$~mJy/beam), compared to the 1.4~GHz data, it
does not reveal as much structural detail as seen at 20~cm. Nevertheless, the
main features, such as the central core and lobes, are still
discernible. 

We use the 1.4~GHz and 324~MHz data to perform a spectral index
analysis.  We find that the region where the central core of CWAT-02
is located is characterized by a flat spectrum ($\alpha \lesssim 0.3$)
likely resulting from the multicomponent emission of the core. The
spectral index steepens to $\alpha \sim 1$ in the area from the core
toward the lobes. In the lobes the spectral index flattens again, 
suggesting possible particle re-acceleration in the hot-spots. The
mean spectral index of the total source is $\alpha=0.6$, which we
adopt for further calculations.

With a total flux density of 15.1~mJy at 1.4~GHz (see Schinnerer et
al. 2010), and 18.2~mJy at 324~MHz (\smo\ et al., in prep), the
observed (not de-boosted) rest-frame 1.4 GHz and 324~MHz luminosities
of CWAT-02 are $1.4 \times 10^{25}$ W Hz$^{-1}$, and $1.7\times
10^{25}$~\wh , respectively. The total radio luminosity, obtained by
integrating the synchrotron spectrum (see e.g.\ eq.~1 in S07) from
10~MHz to 100~GHz is then $L_{\scriptsize\textup{tot}}=2.7 \times
10^{42}$~erg~s$^{-1}$.

\subsection{Doppler boosting and the jet velocity}
\label{sec:boosting}

\begin{figure}
\center{
\includegraphics[bb = 34 370 575 790, width=1.1\columnwidth]{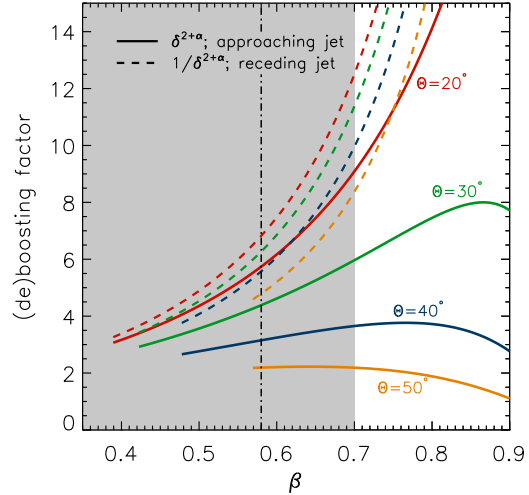}
\caption{ The Doppler boosting factor of the approaching jet (solid
  lines) and the deboosting factor of the receding jet (dashed lines) as
  function of the bulk jet speed $\beta=v/c$ and the jet orientation
  angle to the line of sight $\theta$ for a range of
  jet--to--counter-jet ratios $\geq10$. The value of the speed of
  sound in a relativistic plasma ($0.58c$) is indicated by the
  dash-dotted vertical line. The typical jet speed range ($0.3c -
  0.7c$; adopted from Jetha et al.\ 2006) is designated by the
  gray-scale area. The sound speed sets a lower limit to CWAT-02 jet
  velocity (see text for details).
\label{boost}}
}
\end{figure}

The radio jet luminosity of CWAT-02 is highly asymmetric. A plausible
explanation lies in Doppler boosting which causes the luminosity of
the jet moving towards the observer to be amplified and the luminosity
of the jet moving away to be suppressed, assuming the bulk velocity of
the jet is relativistic.

For a jet moving at a speed of $v=\beta c$, the ratio of the  rest-frame
(Doppler boosted) luminosity $L(\nu)$ to the intrinsic luminosity $L_0(\nu)$ is given 
by (see e.g.\ Lind~\&~Blandford~1985)
\begin{equation}
\frac{L\left(\nu\right)}{L_0\left(\nu\right)}=\left[\gamma \left(1-\beta \cos{\theta}\right)\right]^{-\left(2+\alpha\right)}\equiv \delta^{2+\alpha} \ \mbox{,}
\label{eq:obs2rest}
\end{equation}
where $\gamma=\left(1-\beta^2\right)^{-1/2}$ is the Lorentz factor, $\theta$ 
is the jet orientation angle with respect to the line of sight, and $\alpha$
is the spectral index ($F_{\nu} \propto \nu^{-\alpha}$). The quantity
$$\delta=\left[\gamma\left(1-\beta \cos{\theta}\right)\right]^{-1}$$
is the so-called  Doppler factor.

Applying eq.~(1) to both the approaching and the receding jet and assuming that 
the intrinsic rest-frame luminosities of the two jets are the same, the
ratio of jet to counter-jet fluxes (or monochromatic luminosities) is
obtained as
\begin{equation}
\label{eq:R}
R=\frac{L_{app}}{L_{rec}}=\frac{S_{app}}{S_{rec}}=\left(\frac{1+\beta \cos{\theta}}{1-\beta \cos{\theta}}\right)^{2+\alpha} \ \mbox{.}
\end{equation} 
  
The flux of the southern jet of CWAT-02, assumed to be approaching, out to the bending point, is
$848~\mu$Jy. For the northern jet we can only place an upper limit on
the flux. Taking a $3\sigma$ value times the estimate of the area
over which the jet extends, we obtain the flux of the northern jet to
be $\lesssim~84~\mu$Jy, yielding $R\gtrsim10$. 
With $\alpha=1$ and $R\gtrsim10$, eq.~(2) yields $\beta \cos{\theta}\gtrsim 0.37$.

As only the lower limit on the jet flux ratios is known, and there is
no independent constraint on either the jet velocity or the
orientation angle, it is difficult to break the degeneracy between
$\beta$ and $\theta$. However, we can place limits on the most likely
velocity (and then on $\beta$) as follows. \f{boost} \ shows
the Doppler boosting factor of the approaching jet ($L_{app}/L_0$; 
solid lines) and the deboosting factor of the receding jet
($L_0/L_{rec}$; dashed lines) as a function of the speed of the
jet and the inclination angle, assuming a range of flux ratios
$R\gtrsim10$. For each $\theta$, assuming the same $\beta$ for the
two jets, the expected R is $\sim10$ for a beta value
corresponding to the left limit of the continuous lines (boosting for
the approaching jet) and rapidly increases for higher $\beta$. We
can constrain a possible range of $\beta$ from the observations. 
The first constraint on the jet speed arises from the hot-spots in
CWAT-02.  Assuming the hot-spots are jet-termination shocks (as in FR
II sources), the presence of a hot-spot indicates that the speed of
the jet is higher then the internal speed of sound ($\approx 0.58c$
for a relativistic plasma; see e.g.\ Jetha~et~al.~2006).  This then
sets a likely lower limit on CWAT-02's jet speed of
$\sim0.6c$. Secondly, based on a sample of 30 WATs, combined with
Monte Carlo simulations to infer their bulk jet velocities, Jetha et
al. (2006) found typical WAT jet speeds to be in the range $0.3c -
0.7c$ (indicated by the gray-shaded area in \f{boost} ). Thus,
combining the above two arguments suggests that a likely bulk velocity
of CWAT-02's jets is somewhere in the range $0.6c - 0.7c$. Depending
on the angle between the jets and the line of sight, the observed
luminosity of the approaching jet can then be boosted by a factor ranging
from $\sim2$ (for $\theta\sim50^\circ$) to $\sim9$ (for
$\theta\sim20^\circ$), while the luminosity of the receding jet would
be de-boosted by factors $\sim5$ and $\sim13$ for the same $\theta$
values (see \f{boost} ). Taking the values for the maximum
(de-)boosting ($\theta=20^\circ$; $\beta=0.7$) and assuming that the
intrinsic (non-boosted) luminosities of the jets are equal, we find
that the observed total jet flux (the sum of the approaching and
receding jet flux densities) is a factor of $\sim4.5$ higher than the
intrinsic one. This suggests that the total intrinsic luminosity of
the jets is likely less than a factor of 5 lower compared to the
observed value. However, the observed jet flux (1.77 mJy at 20 cm)
accounts for only 12\% of the total observed flux of CWAT-02 (15.1 mJy). 
Thus, the remaining 88\% arises from non-boosted
features, like the core and the lobes. Therefore, Doppler boosting
affects only a minor portion of the total flux of the source, and the
observed value of the luminosity, reported in Sec. 3.2., is a good
estimate of the total intrinsic luminosity.

\section{The environment of CWAT-02}
\label{sec:hostcl}

\subsection{Voronoi tessellation-based approach: the CWAT-02 host group}
\label{sec:vta}

\begin{figure*}
\center{
\includegraphics[bb=104 340 558 720, scale=0.55]{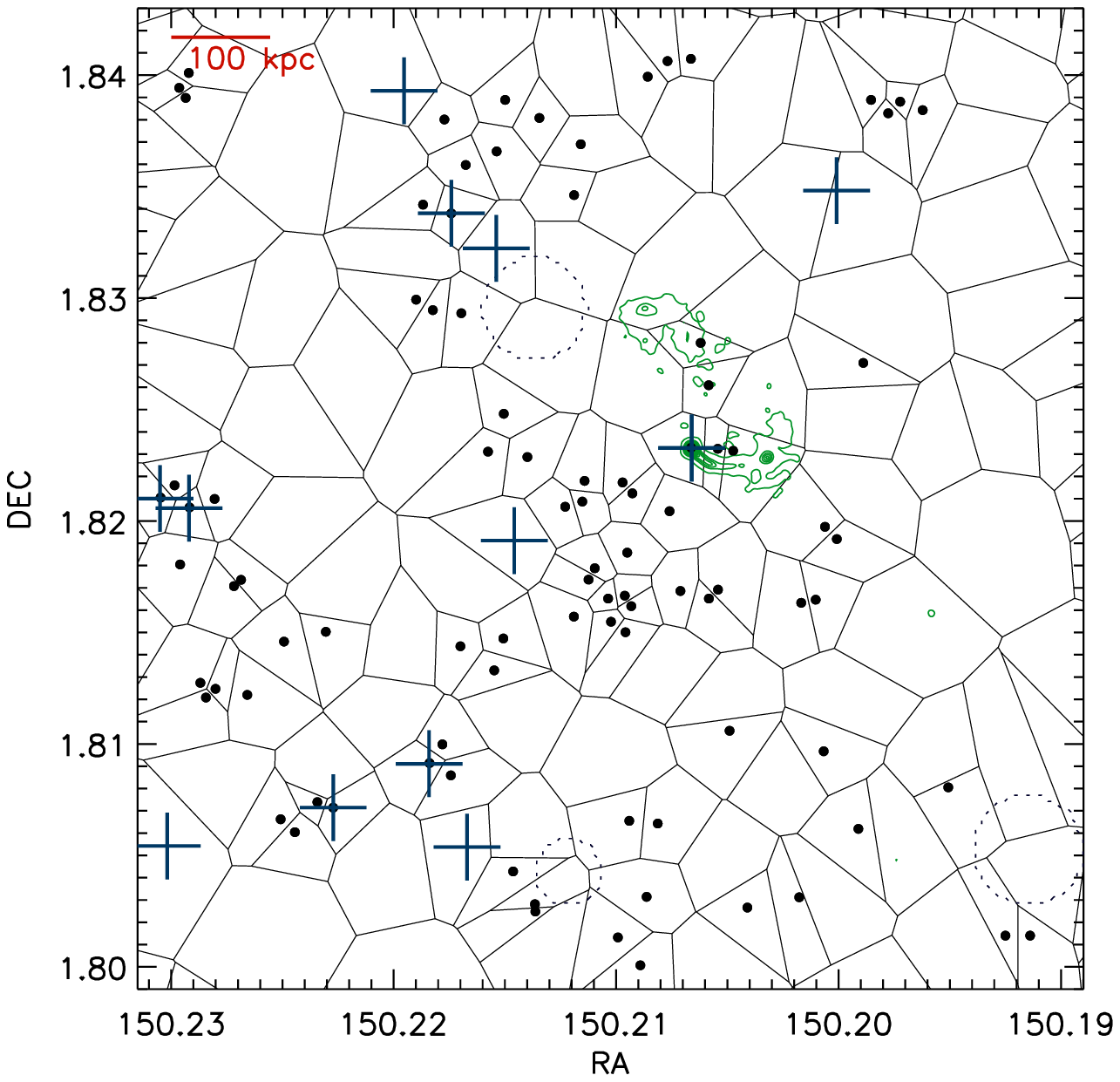} 
\includegraphics[bb=190 340 558 720, scale=0.55]{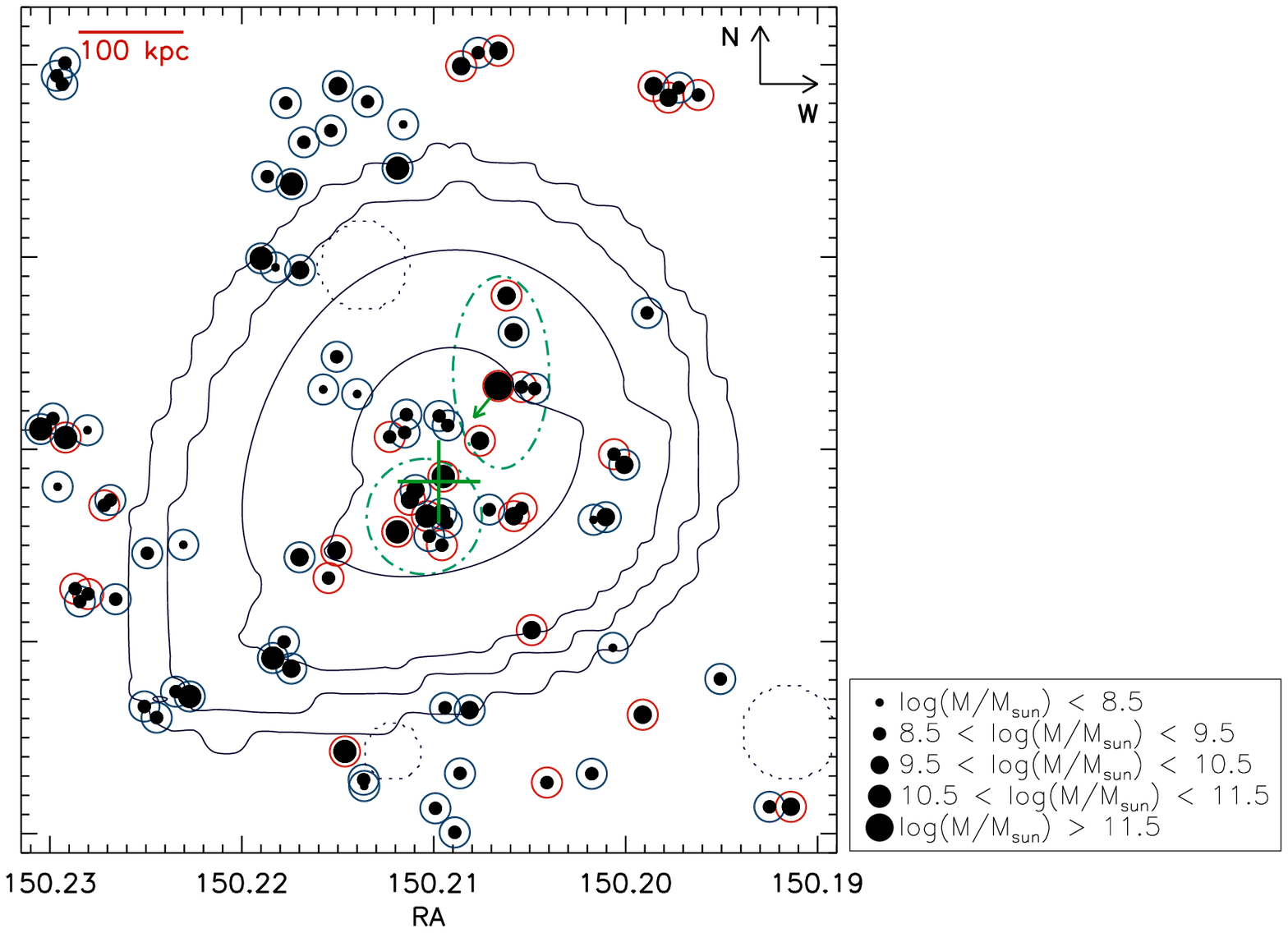} 
\caption{ Left panel: Voronoi tessellation of the area around CWAT-02 
  ($2.64'~\times~2.64'$, corresponding to $\sim 1~\mbox{Mpc} \times 1$~Mpc, 
  which is $\sim 1\%$ of the total area analyzed) 
  overlaid with radio contours. Filled dots represent "high density"
  galaxies that are less than $\sim 500$~kpc away from CWAT-02. Crosses
  indicate the position of the galaxies for which a spectroscopic redshift is
  available. Masked-out regions in the optical, around saturated
  objects, are marked with dotted circles.  The right panel shows the
  same area as in the left panel, but overlaid with X-ray
  contours. The galaxies are represented by encircled dots, where the
  size of each dot is related to the stellar mass of the corresponding
  galaxy, as indicated in the legend. The color of the circle matches
  the color of the galaxy, retrieved from the color-magnitude diagram
  (see \f{fig:cmd} ).  The arrow indicates the velocity direction of
  CWAT-02 (projected onto the plane of the sky; see \f{fig:radio}
  \ ). The cross marks the center of mass of the galaxies within the
  X-ray emission region (see text for details). Dash-dotted lines
  indicate the two assumed merging galaxy subgroups, one is associated
  with the center of mass of the system, and the other contains
  CWAT-02.
\label{fig:vta}
}
}
\end{figure*}

To identify the galaxy overdensity associated with CWAT-02, we use
the Voronoi tessellation-based approach (VTA; e.g. S07; Ramella et 
al. 2001; Kim et al. 2002), a method
particularly favorable for revealing substructure in overdense
environments.  A Voronoi tessellation on a two-dimensional plane
containing positions of galaxies within a fixed redshift range,
divides the plane into convex polygons where each polygon contains one
galaxy.  The inverse of the area of each polygon corresponds then
to the effective local density of the galaxy.  

To determine the clustering in the CWAT-02 region, we select galaxies
that are less than $\sim 13'$ (5~Mpc at $z=0.53$) away from CWAT-02,
and have photometric redshifts within $\Delta z=3\sigma_{\delta
  z/\left(1+z_s\right)}\cdot(1+0.53)$, where $\sigma_{\delta
  z/\left(1+z_s\right)}$ is the typical photometric redshift error
(see \s{sec:intro} ).  To estimate the background density we randomly
distribute the same number of galaxies (11162) over the same area 100
times. We then apply the VTA to each generated field (see Botzler
2004), and calculate the mean local density
$\overline{\rho_{bkg}}$ and its standard deviation $\sigma_{bkg}$ (see 
S07 for more details). All the Voronoi tiles with densities
exceeding the value of $\overline{\rho_{bkg}} +10 \sigma_{bkg} $ are
considered overdense.

\begin{figure}
\center{
\includegraphics[bb = 54 360 575 790, width=1.0\columnwidth]{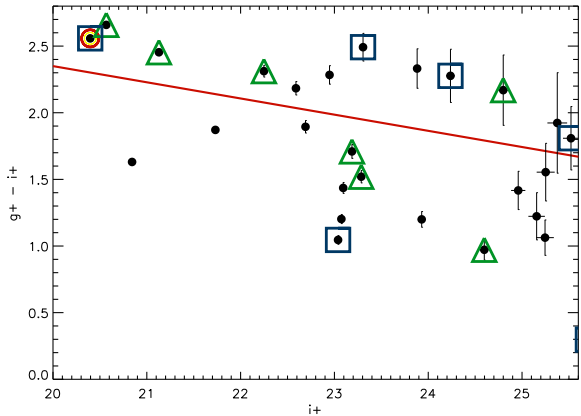}
\caption{ Color-magnitude diagram of the galaxies around CWAT-02. We
  use the indicated line to separate blue from red galaxies.  A
  distinct red sequence is discernible. The triangles indicate the
  members of the subgroup located near the center of the X-ray
  emission, and the squares mark the galaxies forming the subgroup
  dominated by CWAT-02 (see \f{fig:vta} ).  CWAT-02 is indicated by
  the encircled dot.
\label{fig:cmd}}
}
\end{figure}

The spatial distribution of these "overdense" galaxies, shown in
\f{fig:vta} , in the immediate surrounding of CWAT-02 shows an
elongation in the NW-SE direction, with two discernible high-mass
peaks indicated in the right panel of \f{fig:vta} . There is a
strong accumulation of very bright, red and massive early type
galaxies close to the center of the X-ray emission (see
below). However, the most massive and brightest galaxy is CWAT-02's
host galaxy. With a few other galaxies, CWAT-02 makes the northwestern
elongation of the central massive galaxy accumulation.

In \f{fig:cmd} \ we show the color-magnitude diagram of the
"high-density" galaxies (shown in \f{fig:vta} ). The red sequence is
clearly visible. It is interesting that although CWAT-02's host galaxy
is the brightest one in the system, the next three brightest
red sequence galaxies are all concentrated in the clump close to the
center of the X-ray emission, offset from CWAT-02's host by
$\sim19''$, i.e. $\sim120$~kpc at $z=0.53$. This suggests an
un-relaxed state of the cluster and will be further discussed in
\s{sec:discussion} .

The existence of a galaxy agglomeration around CWAT-02 is verified by
12 spectroscopic redshifts ($z$ in the range from 0.5242 to 0.5341, 
with an average redshift of 0.5291) present in the 
region of interest (see \f{fig:vta} \ \& Tab.~1). Furthermore, the entire
structure is embedded in the densest large-scale structure (LSS)
component at $z=0.53$ (Scoville~et~al.~2007; in prep.). It is located
at the north-western outskirts of this LSS component which is fairly
elongated, and extends over $\sim 5'$ ($\sim 2$~Mpc at $z=0.53$) in
the NW-SE direction.

\begin{table}[!h]
\caption{Spectroscopic redshifts for the galaxies around CWAT-02, drawn from zCOSMOS (Lilly et al. 2007, 2009) and SDSS (Abazajian et al. 2009).}
\begin{center}
\begin{tabular}{|c|c|c|c|}
\hline
RA & DEC & origin & $z_\mathrm{spec}$ \\
\hline \hline
150.20009   & 1.83482   & zCOSMOS & 0.5296 \\ %& 2.5 \\
150.20661   & 1.82327   & SDSS & 0.5302 \\ % & 4.0 \\
150.21458   & 1.81912   & zCOSMOS & 0.5289 \\ % & 3.5 \\
150.21539   & 1.83223   & zCOSMOS & 0.5341 \\ % & 4.5 \\
150.21670   & 1.80537   & zCOSMOS & 0.5294 \\ % & 2.5 \\
150.21741   & 1.83380   & zCOSMOS & 0.5314 \\ % & 3.5 \\
150.21841   & 1.80911   & zCOSMOS & 0.5280 \\ % & 3.5 \\
150.21953   & 1.83930   & zCOSMOS & 0.5280 \\ % & 1.5 \\
150.22272   & 1.80714   & zCOSMOS & 0.5300 \\ % & 3.5 \\
150.22920   & 1.82058   & zCOSMOS & 0.5242 \\ % & 4.5 \\
150.23018   & 1.80542   & zCOSMOS & 0.5301 \\ % & 4.5 \\
150.23050   & 1.82101   & zCOSMOS & 0.5250 \\ % & 3.5 \\
\hline
\end{tabular}
\end{center}
\end{table}

\subsection{X-ray properties of the galaxy group}

The extended X-ray emission from the diffuse hot gas within the galaxy
group hosting CWAT-02 is shown in \f{fig:vta} \ (right panel). As
reported in the COSMOS galaxy cluster catalog (Finoguenov et
al.\ 2007; in prep), the total X-ray luminosity of the ICM, in the
$0.1 - 2.4$~keV band, is $L_X=(6\pm
1)\times10^{42}~\mbox{erg~s}^{-1}$, the total mass embedded within
$r_{200}$ is $(5.2\pm0.4)\times10^{13}~\mathrm{M}_\odot$, and the
temperature is $kT=1.04\pm0.06$~keV, thus consistent with properties
of a galaxy group. The total mass and temperature have been estimated
using the scaling relations from Leauthaud et al. (2010).  The group's
extended X-ray emission has been identified at high significance. We
note that two strong X-ray point sources have been subtracted from the
group's emission,  (NW, and SE from CWAT-02), thus potentially
biasing the X-ray group center determination.  To constrain the group
center more robustly, we compute the center of mass using the
potential group member galaxies within the region of X-ray emission
(see \s{sec:vta} ). The center of mass (see \f{fig:vta} ) roughly
coincides with the agglomeration of red, massive galaxies SE of
CWAT-02, and is very close to the peak of the X-ray emission.

In addition to the group's extended X-ray emission, CWAT-02 itself is 
detected as a faint source in the Chandra data (identified as Chandra source
\#1292; Civiano et al., in prep.), with a flux of 
$4.5\times10^{-16}$~erg~cm$^{-2}$~s$^{-1}$ in the soft band (0.5 - 2.0 keV; 
Elvis et al. 2009).

\section{Discussion}
\label{sec:discussion}

\subsection{WATs as tracers of dynamically young clusters}
\label{sec:clust}

The existence of CWAT-02's host galaxy group at $z=0.53$ has been
independently determined in three different wavelength windows (radio,
optical, and X-ray).  We find that although it is the brightest and
most massive galaxy in the analyzed region, CWAT-02 is not, as one
would expect in a relaxed system, at the center of its host group, but
offset from it to the NW. The next three brightest massive galaxies
are located in the agglomeration closest to the center of mass of the
group. This is an indication of a disturbed system, that is probably
undergoing a process of group or sub-group merger. To test this idea in the
following we put constraints on the velocity of CWAT-02's host galaxy
relative to the ICM that is needed to explain the observed bending of
the radio jets, and compare it to velocities expected in relaxed
systems.

If the bending of the radio jet is generated by ram pressure, then a
non-negligible motion of the galaxy relative to the ICM must
exist. To shed light on this, we make use of the model developed by
Begelman, Rees \& Blandford (1979), which assumes that the ram
pressure exerted on the galaxy, as it moves through the ICM, is
balanced by the centrifugal force exerted by the jet as the jet moves with
a bulk velocity outwards from the host galaxy and thereby curves. We make 
use of the Euler equation, written in the form
\begin{equation}
\frac{\rho_{ICM}\,v_{gal/ICM}^2}{h}=\frac{\rho_j\,v_j^2}{R_j} \ \mbox{,}
\label{eq:euler}
\end{equation}
where $v_j$ is the bulk jet speed, $h$ is the jet scale height, $R_j$
is the radius of curvature of the jet, $v_{gal/ICM}$ is the galaxy
velocity relative to the ICM and $\rho_\mathrm{ICM}$ and $\rho_j$ are
the ICM and jet densities, respectively. Although this model presumes
a non-relativistic jet velocity, it is a reasonable approximation in
case of a light and moderately relativistic jet (see
Hardcastle~et~al.\ 2005; Jetha~et~al. 2006).  For CWAT-02 we estimate
a scale height of $h \sim 2.2"$ ($\sim 14~$kpc at $z=0.53$) and a
radius of curvature of $R_j \sim 9.0"$ ($\sim 57~$kpc at
$z=0.53$). Note, however, that the jet and ICM densities for the
CWAT-02 system are not well constrained. For this reason in
\f{fig:vgal} \ we show the dependence of $v_{gal/ICM}$ on the jet to
ICM density ratio, $\rho_j/\rho_\mathrm{ICM}$. In general, typical
$\rho_j/\rho_\mathrm{ICM}$ values used in hydrodynamical simulations
of radio jets are of the order of $10^{-4}-10^{-2}$
(e.g.\ \citealt{rossi04}). Thus, for the lower limit value of
$\rho_j/\rho_\mathrm{ICM}=10^{-4}$ we obtain a velocity of
$v_{gal/ICM}\sim900$~\kms . Note that even for density ratios lower
than this value $v_{gal/ICM}$ stays significant, i.e.\ more than several hundred \kms . 

\begin{figure}
\center{
\includegraphics[bb = 90 440 490 720, width=1.0\columnwidth]{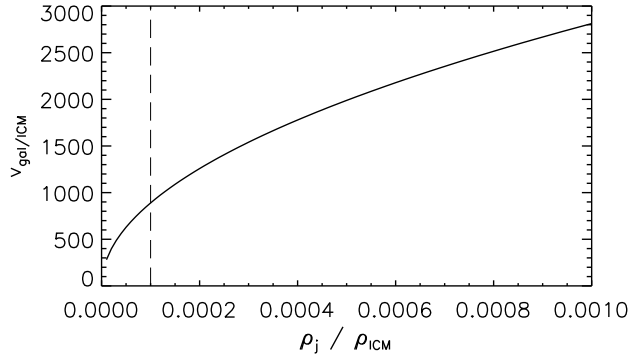}
\caption{ The velocity of CWAT-02's host galaxy relative to the ICM
  (estimated via eq.~\ref{eq:euler}) as a function of the jet to ICM
  density ratio, $\rho_j/\rho_\mathrm{ICM}$ (thick line). The dashed
  line shows a low $\rho_j/\rho_\mathrm{ICM}$ limit, typically
  assumed in hydrodynamical simulations (see text for more details).
\label{fig:vgal}}
}
\end{figure}

Based on our spectroscopic data we can further put limits on the
peculiar line-of-sight velocity of CWAT-02's host galaxy. Note,
however, that this is based on the assumption that the galaxies in the
core of the group (surrounding CWAT-02) have a similar velocity
distribution as those with spectroscopic redshifts, that are located
predominantly at the outskirts of the system.  With this in mind, for
the 12 galaxies from Tab.~1 we compute the line-of-sight velocities as
$v=z\cdot c$, where $v$ is the velocity, $z$ is the redshift, and $c$
is the speed of light.  To estimate the median velocity and the
velocity spread for CWAT-02's group we use biweight statistics, that
was shown to be superior for samples containing $\sim10$ galaxies
\citep{beers90}. The redshift distribution is shown in \f{fig:vdisp} .
 The obtained biweight velocity dispersion is $437$~\kms. Note that
based on the X-ray luminosity of the group, a velocity dispersion of
$300-500$~\kms\ is expected \citep{mulchaey00}.

The velocity difference between CWAT-02's host galaxy and the biweight 
mean velocity of all the group galaxies is $244$~\kms.
This is higher than the expected line-of-sight peculiar velocities
of $\lesssim150$~\kms\ of brightest group galaxies located in relaxed
systems (Beers et al.\ 1995; see also Oegerle \& Hill 2001). Thus it
independently suggests a disturbed state of the system. The lower limit 
of the relative velocity between the CWAT-02's host galaxy and the ICM, in
the projected plane of the sky, of $\sim900$~\kms\ is noticeably higher
then the derived peculiar line-of-sight velocity of the galaxy of 244~\kms.
Such a discrepancy suggests that a significant fraction of the relative 
velocity between the CWAT-02's host galaxy and the ICM likely arises 
from the bulk motion of the ICM itself. Large bulk velocities of the ICM 
can be caused by group mergers (Roettiger et al.\ 1996; Sato et al.\ 2008) 
that can also explain the elongation of the X-ray emission that is observed 
in CWAT-02's host group (Roettiger et al.\ 1996).

\begin{figure}
\center{
\includegraphics[bb = 90 450 490 720, width=1.0\columnwidth]{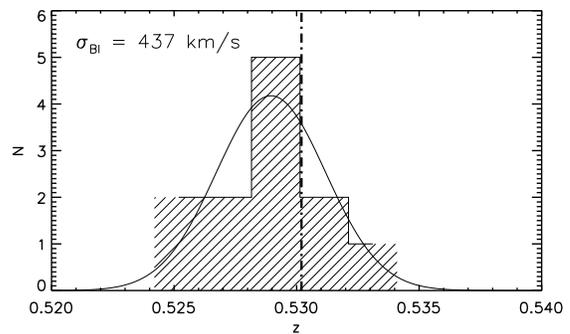}
\caption{ The redshift distribution of the 12 galaxies in the CWAT-02
  region with spectroscopic redshifts (see Tab.~1; dashed
  histogram). Biweight statistics has been used to infer the spread 
  $\sigma_\mathrm{BI}$ for the system (indicated in the panel). 
  The dash-dotted line corresponds to the velocity of CWAT-02's host 
  galaxy, offset from the mean velocity by $244$~\kms\ (see text for details).
\label{fig:vdisp}}
}
\end{figure}

On large scales the CWAT-02 host group is part of the strongest LSS
component at $z=0.53$ (Scoville et al.\ 2007; in prep), which extends over
2~Mpc in the NW-SE direction, coincident with the elongation of the
X-ray and optical distributions, as well as the motion of CWAT-02,
relative to the ICM (projected on the plane of the sky). Given the
location of the group at the NW outskirts of the LSS component, it is
plausible to expect that the entire group will eventually migrate
towards the center of the LSS component, and form a massive relaxed
cluster.

Our results also show that wide-angle tail galaxies are efficient
tracers of large-scale structure over-densities. This is further
strengthened by an analysis of the large-scale environment of  
$\sim10$ other WATs in the COSMOS field (see Schinnerer et al.\ 2007) that
yields that all (but one\footnote{The galaxy has a C-shaped radio
morphology. Its optical counterpart is a type 1 AGN at $z=1.35$, and
the radio jets are strongly Doppler boosted. Thus, the apparent
bending of the jets may possibly arise from relativistic effects
rather than the motion of the galaxy though a dense medium.} )  are 
associated with LSS overdensity peaks typically extending over 2~Mpc 
(see also S07).

In summary, our results support the hypothesis that wide-angle tail
galaxies inhabit gravitationally disturbed systems in the 
process of formation. This is consistent with previous results 
(e.g. Pinkney et al. 1994, Loken et al. 1995, Gomez et al. 1997, 
Sakelliou et al. 1996, Sakelliou \& Merrifield 2000)
and demonstrates the usefulness of WATs as tracers of dynamically 
young groups and clusters. In the following analysis we demonstrate 
the value of identifying merging systems, in the context of the 
missing baryon problem in galaxy groups.

\subsection{Can radio-AGN heating expel baryons from CWAT-02's host group?}

The interest in radio galaxies has been renewed in the last years as
radio-AGN heating is regularly invoked in cosmological models as a key
ingredient to explain the observed galaxy and galaxy cluster/group
properties (Croton et al.\ 2006; Bower et al.\ 2006; Bower et
al.\ 2008). Recent cosmological models assume that radio-AGN heating
is powerful enough to expel a fraction of baryons from the
cluster/group potential well (Bower et al.\ 2008). Such a process
provides a satisfactory solution for the missing baryon problem on
galaxy group scales, i.e.\ it can potentially reconcile the observed
discrepancy between the baryon to dark matter mass ratio in galaxy
groups and the WMAP-CMB value (see Giodini et al.\ 2009; 2010 and
references therein). The first observational support for this scenario
has been presented by Giodini et al.\ (2010). For each of the 16
X-ray selected galaxy groups in the COSMOS field that host a radio
galaxy Giodini et al.\ have computed the output energy that the radio
galaxy exerts onto its environment (over the host galaxy's lifetime),
as well as the binding energy of the ICM. Comparing these two for all
systems, they find that the energetics of radio AGN, computed in this
way, may account for expelling gas from the group's potential well.
One of the 16 systems analyzed by Giodini et al. (2010; identified as
XID35) is CWAT-02's
host group for which they infer a mechanical radio output energy of
$0.2-9.9\times10^{61}$~erg (the range is a result of the 0.85~dex
scatter in the scaling relations used to obtain this result; see
\citealt{birzan08}). They find a comparable X-ray binding energy of
$2.4-4\times10^{61}$~erg, suggesting that the radio-AGN heating done
by CWAT-02 is powerful enough to expel gas from its host group.  We
discuss below how the findings based on the in-depth analysis of the
CWAT-02 host system presented here affect this result.

Our Doppler boosting analysis, presented in \s{sec:boosting} , has
shown that the relativistic boosting does not significantly affect the
total intrinsic radio luminosity of CWAT-02, which was used to compute
the output energy by Giodini et al\ (2010). Therefore, Doppler
boosting of the jets of CWAT-02 does not substantially alter the above
given radio output energy. 
However, it is possible that massive galaxy members in CWAT-02's
host group, that are currently radio-silent, have experienced
radio-AGN phases in their past and thereby contributed to the energy
budget required to expel baryons from the group. In general,
radio-emission in central massive galaxies located in galaxy clusters
or groups is thought to originate from a self-regulating process (Best
et al.\ 2005; Hardcastle et al.\ 2007, Merloni et al.\ 2008, Tasse et
al.\ 2009; S09). In this picture radio activity is induced via cooling
of hot gas onto the galaxy (Hardcastle et al.\ 2007, Merloni et al.\
2008). In turn, radio outflows heat the surrounding gas thus limiting
its inflow and terminating the source of radio emission.  The
brightest, reddest and most massive galaxies in CWAT-02's host group
are dispersed in two clumps (see \f{fig:vta} ) which is likely a
result of group merger (as discussed in the previous section). Hence,
it is possible that (some of) the most prominent galaxies in the
merging system were once the central galaxies in the merging
constituents. As such, based on the above outlined scenario, in these
galaxies re-occurring radio outflows (that couple with the ICM) would
be expected (Fabian 1994; Birzan et al.\ 2004, 2008, Giodini et al.\
2009; 2010). While not much change in radio power of such massive
galaxies is (on average) expected with cosmic time (see \smo\ et al.\
2009), the binding energy of the merging constituents could have been
substantially lower (compared to the merged system), thus
facilitating the removal of gas.  In summary, the radio output energy
budget in CWAT-02's host group may be powerful enough to expel baryons
from its host group, strengthening the results of Giodini et al.\
(2010) who showed that radio-AGN heating may (on average) account for
the missing baryons in galaxy groups.

\section{Summary}

We have presented an analysis of a wide-angle tail galaxy in the
COSMOS field (CWAT-02) and its host galaxy group at $z=0.53$. The
in-depth study of the galaxy's environment, morphology and velocity
suggests an unrelaxed state of the host group, possibly caused by a
galaxy group merger. This is consistent with the idea that WAT
galaxies can be used as good tracers of dynamically young, unrelaxed
systems. The analysis of radio energy outflows from CWAT-02 suggests
that the outflows (over the host galaxy's lifetime) may be powerful
enough to expel gas from the group.

\acknowledgments AO thanks California Institute of Technology for
generous support through NASA grants 1292462 and 1344606.  VS
acknowledges support from the Owens Valley Radio Observatory, which is
supported by the National Science Foundation through grant
AST-0838260. The research leading to these results has received funding 
from the European Community's Seventh Framework Programme (FP7/2007-2013) 
under grant agreement n° No 229517. VS \& AO thank Unity through Knowledge Fund (www.ukf.hr)
for collaboration support through the 'Homeland Visit' grant. 
SG acknowledges support by the DFG Cluster of Excellence "Origin 
and Structure of the Universe" (www.universe-cluster.de).
CC acknowledges support from the Max-Planck Society and the Alexander von
Humboldt Foundation through the Max-Planck-Forschungspreis 2005. GZ
acknowledges support from an INAF contract PRIN-2007/1.06.10.08 and an
ASI grant ASI/COFIS/WP3110 I/026/07/0.

{}

%\newpage

%\newpage

\end{document}